\documentclass{ws-mpla}
\usepackage{cite}
\usepackage{graphicx}
\usepackage{amsmath}
\usepackage{amssymb}
\usepackage{epsfig}
\usepackage{euscript}
\usepackage{fancybox}
\usepackage{color}

\def\mathswitchr#1{\relax\ifmmode{\mathrm{#1}}\else$\mathrm{#1}$\fi}

\newcommand {\pslash}{\hbox{$\not\hbox{\kern-2.3pt $p$}$}}

\usepackage{epic}

\def\alf1{ {\alpha\over\pi} }

\begin{document}
\markboth{B.F.L. Ward}
{RUNNING OF THE COSMOLOGICAL CONSTANT AND ESTIMATE OF ITS VALUE IN QUANTUM GENERAL RELATIVITY}
\catchline{}{}{}{}{}
 
\title
{\rm \normalsize RUNNING OF THE COSMOLOGICAL CONSTANT AND ESTIMATE OF ITS VALUE IN QUANTUM GENERAL RELATIVITY}

\author{\footnotesize\rm   B.F.L. WARD}

\address{\footnotesize\rm Department of Physics, One Bear Place \# 97316,
 Baylor University,\\ 
Waco, Texas, 76798-7316, USA\\
bfl\_ward@baylor.edu}
\maketitle

\pub{Received (Day Month Year)}{Revised (Day Month Year)}

 \begin{abstract}
We present the connection between the running of the cosmological
constant and the estimate of its value in the resummed quantum gravity 
realization of quantum general relativity. We also address in this way
some of the questions that have been raised concerning 
this latter generalization and application 
of the original prescription of Feynman
for the formulation of quantum general relativity.

\end{abstract} 

\ccode{PACS Nos.: 04.60.Bc;04.62.+v;11.15.Tk}
\centerline{\footnotesize\rm Contributed paper to the Special Issue: ``Fundamental Constants in Physics and Their Time Variation''}
\centerline{\footnotesize\rm (Modern Physics Letters A, Guest Ed. Joan Sol{\`a})}
\centerline{\footnotesize\bf BU-HEPP-14-08, Nov., 2014}

%
\renewcommand{\baselinestretch}{0.1}
\footnoterule


\def\Kmax{K_{\rm max}}\def\ieps{{i\epsilon}}\def\rQCD{{\rm QCD}}
\renewcommand{\theequation}{\arabic{equation}}
\renewcommand\thepage{}
\parskip.1truein\parindent=20pt\pagenumbering{arabic}\par
\section{\bf Introduction}\label{intro}\par
As one can see in Refs.~\cite{attk1,sola-shp1,sola-shp2,sola-shp3,sola-shp4,sola-shp5,sola-shp6,sola-shp7,b-s} there has been some controversy about
the meaning of a running cosmological constant in quantum field theory.
In sum in Ref.~\cite{attk1}, the invariance of the physical vacuum energy density under renormalization group action is used to argue  
that the total response
of this quantity to a change in the renormalization scale, $\mu$, is 
zero, so that it does not actually run. The authors in Ref.~\cite{sola-shp1} argue that, while the total response of the vacuum energy density
to a change in such a
scale is zero, 
this still allows for that part of Einstein's theory that we ``see'' 
at low energy to contribute to the implicitly running part 
of the vacuum energy density, which is then
compensated by the dependence on the running scale 
due to both known contributions and unknown contributions from  
the possible UV completion 
of Einstein's theory.
Here, we will present arguments that generally agree with this latter view
and with that in Refs.~\cite{b-s,mrkkn}, where we use a UV finite approach~\cite{bw1,bw2,bw3,bw4,bw5} to
quantum general relativity developed from an extension of
Feynman's formulation~\cite{rpf1,rpf2} of Einstein's theory.
This we do in the next Section
\footnote{We need to stress that the arguments that are given in Ref.~\cite{attk1}
do not disagree with those we present here, as we have explained in Ref.~\cite{runcos1}, when ``apples'' are compared to ``apples''.}.  
\par
Having done this, we then show in Section 3 how the running
of the cosmological constant and the Newton constant 
are featured in a first principles estimate of the observed value of
the cosmological constant in the Planck scale cosmology scenario of 
Refs.~\cite{bon-reut2}. Section 4 contains our  summary remarks.\par

\section{Review of the Running of the Cosmological Constant}

We begin with a recapitulation of the arguments in Ref.~\cite{runcos1}.
In this connection, there is one
important point of definition. Our discusion will be based 
on Einstein's equation
\begin{equation}
R_{\mu\nu}-\frac{1}{2}Rg_{\mu\nu} + \Lambda g_{\mu\nu}=-8\pi G_N T_{\mu\nu}
\label{ein1}
\end{equation}
where $R_{\mu\nu}$ is the contracted Riemann tensor, $R$ is the curvature scalar, $g_{\mu\nu}$ is the metric of space-time, $G_N$ is Newton's constant,
$T_{\mu\nu}$ is the matter energy-momentum tensor and $\Lambda$ is the
cosmological constant as we will define it in our discussion. 
We follow Feynman
and expand about flat Minkowski space with the metric representation
\begin{equation}
g_{\mu\nu}=\eta_{\mu\nu}+2\kappa h_{\mu\nu}
\end{equation}
where $\kappa=\sqrt{8\pi G_N}$, $\eta_{\mu\nu}=\text{diag}(1,-1,-1,-1)$,
so that $h_{\mu\nu}$ is the quantum fluctuating field of the graviton here.
If we take the vacuum expectation value of (\ref{ein1}), 
we can move the purely gravitational contribution to the VEV,
which arises from the nonlinear part of the
``geometric'' side of Einstein's equation, to the right-hand side to get
\begin{equation}
\Lambda \eta_{\mu\nu}=-8\pi G_N <0|t_{\mu\nu}|0>
\label{cos1}
\end{equation}
where now
\begin{equation}
<0|t_{\mu\nu}|0>=<0|\left(T_{\mu\nu}+\frac{1}{\kappa^2}(R_{\mu\nu}-\frac{1}{2}Rg_{\mu\nu})|_{\text{nonlinear}}\right)|0>.
\label{cos1a}
\end{equation} 
From (\ref{cos1}), we see that, to any finite order in $\kappa$, as the 
tensor $t_{\mu\nu}$ inside the VEV operation
in (\ref{cos1a}) is conserved in the ``flat space'' sense~\cite{wein1}, 
it has zero anomalous dimension
and this proves that $\Lambda$ runs because $G_N$ runs. That $G_N$ runs can be inferred 
immediately from the Dyson resummation
of the graviton propagator and the conservation of $t_{\mu\nu}$: in complete
analogy with QED, the ``invariant charge'' then obtains
\begin{equation}
\kappa^2(q^2)= \frac{\kappa^2}{1+\kappa^2\Pi(q^2,\mu^2,\kappa^2)}
\end{equation}
when $\kappa$ is renormalized at the point $\mu^2$ and $\Pi(q^2,\mu^2,\kappa^2)$is the respective transverse traceless renormalized proper 
graviton self-energy function.\par
The authors in Ref.~\cite{attk1} seem to identify the discussion of the
running of the cosmological constant $\Lambda$ and the discussion
of the running of the vacuum
energy density. From the definition in Einstein's equation (\ref{ein1}), the two
attendant quantities
are in fact related by a factor of $-8\pi G_N$. Observe that, as
the arguments in Ref~\cite{attk1} would 
show as well that the vacuum energy density
does not run, it again would follow that $\Lambda$ as defined here
runs with $G_N$.\par
We now agree to call the object analyzed in Ref.~\cite{attk1} by its
proper name the vacuum energy density $V_{\text{vac}}$. 
From (\ref{cos1}), the relation
\begin{equation}
\Lambda =-8\pi G_N V_{\text{vac}}
\label{cos2}
\end{equation}
holds. The arguments in Ref.~\cite{sola-shp1,b-s} also
seem to be
concerned with $V_{\text{vac}}$ rather than with $\Lambda$
as defined in (\ref{ein1}). Accordingly, let us now 
comment further with emphasis regarding the running physics
of $V_{\text{vac}}$.\par
Specifically, we are working with the entire set of degrees of freedom
in Einstein's theory as formulated by Feynman, 
that is to say, we are working with
the entire set of degrees of freedom in $h_{\mu\nu}$ for example.
By isolating the physics on a given scale
Wilson has shown~\cite{wil1,wil2} that it is possible to 
formulate the solution of the theory in the form of scale transformations
which evolve the theory from one scale to the next, Wilsonian renormalization
group transformations.  
In Refs.~\cite{bon-reut1,bon-reut2,bon-reut3,bon-reut4,bon-reut5,bon-reut6,bon-reut7a,bon-reut7b,bon-reut7c,bon-reut7d,bon-reut8,bon-reut9} Wilson's approach has been pursued
in realizing Weinberg's asymptotic safety approach~\cite{wein2} 
to Einstein's theory.
If we thin the degrees of freedom a la Wilson to those relevant to a 
given scale $\mu$, as it is done in Refs.~\cite{bon-reut1,bon-reut2,bon-reut3,bon-reut4,bon-reut5,bon-reut6,bon-reut7a,bon-reut7b,bon-reut7c,bon-reut7d,bon-reut8,bon-reut9} for example,
the effective action at this scale will give the equation
such as Einstein's (\ref{ein1}), with perhaps some 
higher dimensional operators added for a given level of accuracy,
but for the theory with the thinned degrees
of freedom relevant for the scale $\mu$ 
, with the attendant effective couplings at the scale $\mu$,
and this will mean that only that part
of the vacuum energy density relevant to the physics
on the scale $\mu$ will
enter into relations such as (\ref{cos1}), (\ref{cos2}). This 
means that we have
\begin{equation}
\Lambda(\mu) =-8\pi G_N(\mu) V_{\text{vac}}(\mu)
\label{cos3}
\end{equation}
following the general development of Wilson's renormalization group theory
and dropping possible irrelevant operator terms.
We conclude that both $\Lambda$ and $V_{\text{vac}}$ run when
the theory is solved a la Wilson. This is borne out by the results
in Refs.~\cite{bon-reut1,bon-reut2,bon-reut3,bon-reut4,bon-reut5,bon-reut6,bon-reut7a,bon-reut7b,bon-reut7c,bon-reut7d,bon-reut8,bon-reut9}. It also supports the 
arguments in Refs.~\cite{sola-shp1,sola-shp2,sola-shp3,sola-shp4,sola-shp5,sola-shp6,sola-shp7,b-s,mrkkn}.\par
The basic physics underlying this running of 
both $\Lambda$ and $V_{\text{vac}}$ is as follows. If we do not thin the
degrees of freedom, we can identify a scale~\cite{sola-shp1,sola-shp2,sola-shp3,sola-shp4,sola-shp5,sola-shp6,sola-shp7} $\mu=0$ parameter
that corresponds to the vacuum energy density of the universe
at arbitrarily long wavelengths and call that $V_{\text{vac}-\text{phys}}$
and we can then use Einstein's equation to identify $\Lambda_{\text{phys}}=-8\pi G_N(0)V_{\text{vac}-\text{phys}}$, where $G_N(0)$ is Newton's constant
at zero momentum transfer. These quantities 
$V_{\text{vac}-\text{phys}},\;\Lambda_{\text{phys}}$
would then be invariant under renormalization
and they would not run with changes in the renormalization scale $\mu$.
These are the quantities that the authors in Refs.~\cite{attk1} would appear to have in mind when they assert that {\em the} vacuum energy does not run.
On the other hand, following what is done in Refs.~\cite{bon-reut1,bon-reut2,bon-reut3,bon-reut4,bon-reut5,bon-reut6,bon-reut7a,bon-reut7b,bon-reut7c,bon-reut7d,bon-reut8,bon-reut9} or following
Refs.~\cite{pol1,pol2} and implementing Wilsonian renormalization group theory,
we are naturally led to effective actions in which degrees of freedom have been thinned on a scale $\mu$ and the corresponding values of
$\Lambda$ and $V_{\text{vac}}$,~$\Lambda(\mu),\; V_{\text{vac}}(\mu)$, respectively, for the attendant effective action will
run with $\mu$. If a degree of freedom is integrated out of the path
integral for the theory, all of its quanta are replaced by their effects
on the remaining degrees of freedom. 
This means 
that in general $V_{\text{vac}}$ runs. We turn next to the implications
of such running in estimating the observed value of the cosmological constant
as we have done in Ref.~\cite{darkuni}.
\par
\section{Interplay of Running Parameters and Estimating the Observed Value of Lambda}
One primary motivation for considering the running of the parameters
in the Einstein-Hilbert theory is 
Weinberg's suggestion~\cite{wein2} that the theory may be asymptotically safe, with an S-matrix
that depends on only a finite number of observable parameters, due to
the presence of a non-trivial UV fixed point, with a finite dimensional critical surface
in the UV limit. This suggestion has received significant support from the calculations in Refs.~\cite{bon-reut3,bon-reut4,bon-reut5,bon-reut6,bon-reut7a,bon-reut7b,bon-reut7c,bon-reut7d,bon-reut8,bon-reut9}.
Using Wilsonian~\cite{wil1,wil2,pol1,kgw1,kgw2,kgw3} field-space exact renormalization 
group methods, the latter authors obtain results which support the existence of Weinberg's  UV fixed-point for the Einstein-Hilbert theory. 
Independently, we have shown~\cite{bw1,bw2,bw3,bw4,bw5} that the extension of the amplitude-based, exact resummation theory of Ref.~\cite{yfsa,yfsb,jad-wrd1,jad-wrd2,jad-wrd3,jad-wrd4,jad-wrd5,jad-wrd6,jad-wrd7,jad-wrd8,jad-wrd9,jad-wrd10,jad-wrd11,jad-wrd12,jad-wrd13,jad-wrd14,jad-wrd15,jad-wrd16,jad-wrd17} to the Einstein-Hilbert theory leads to UV-fixed-point behavior for the dimensionless
gravitational and cosmological constants. We have called the attendant resummed theory, which is actually UV finite,
resummed quantum gravity. 
We note that causal dynamical triangulated lattice methods have been used in Ref.~\cite{ambj} to show more evidence for Weinberg's asymptotic safety behavior\footnote{The model in Ref.~\cite{horva} realizes many aspects
of the effective field theory implied by the anomalous dimension of 2 at the Weinberg UV-fixed point but it does so at the expense of violating Lorentz invariance.}.
One can view our results in Refs.~\cite{bw1,bw2,bw3,bw4,bw5} as helping to put the results in Refs.~\cite{bon-reut3,bon-reut4,bon-reut5,bon-reut6,bon-reut7a,bon-reut7b,bon-reut7c,bon-reut7d,bon-reut8,bon-reut9,ambj} on a more firm theoretical foundation insofar as issues of cut-offs/gauge or lattice 
artifacts do not 
arise in our calculations.
\par
Continuing from this latter perspective, we observe that
the attendant phenomenological
asymptotic safety approach in 
Refs.~\cite{bon-reut3,bon-reut4,bon-reut5,bon-reut6,bon-reut7a,bon-reut7b,bon-reut7c,bon-reut7d,bon-reut8,bon-reut9} 
to quantum gravity has been applied
in Refs.~\cite{bon-reut1,bon-reut2} 
to provide an inflatonless realization\footnote{The authors in Ref.~\cite{sola1} also proposed the attendant 
choice of the scale $k\sim 1/t$ used in Refs.~\cite{bon-reut1,bon-reut2}.} of the successful
inflationary model~\cite{guth1,guth2,linde} of cosmology
: the standard Friedmann-Walker-Robertson classical descriptions 
are joined smoothly onto Planck scale cosmology developed from the attendant UV fixed point solution. In this way
a quantum mechanical 
solution is obtained to the horizon, flatness, entropy
and scale free spectrum problems. Using the new
resummed theory~\cite{bw1,bw2,bw5} of quantum gravity, 
the properties as used in Refs.~\cite{bon-reut1,bon-reut2} 
for the UV fixed point of quantum gravity are reproduced in Ref.~\cite{bw4} with the bonus of 
``first principles''
predictions for the fixed point values of
the respective dimensionless gravitational and cosmological constants. 
In what follows, we show how the analysis in Ref.~\cite{bw4} 
can be carried forward~\cite{darkuni} to
an estimate
for the observed cosmological constant $\Lambda$ in the
context of the Planck scale cosmology of Refs.~\cite{bon-reut1,bon-reut2}.
The quantum field theoretic running of parameters 
a la Wilson will be seen to play an essential role in the estimate.
We comment on the reliability and possible implications of the result, 
as the estimate will be seen
already to be relatively close to the observed value~\cite{cosm1a,cosm1b,pdg2008}.  
The closeness of our estimate to the experimental value again gives, at the least, some more credibility to the new resummed theory as well as to the methods in Refs.~\cite{bon-reut3,bon-reut4,bon-reut5,bon-reut6,bon-reut7a,bon-reut7b,bon-reut7c,bon-reut7d,bon-reut8,bon-reut9,ambj}\footnote{We do want to caution against overdoing this closeness to the experimental value.}.
\par
We present the remaining discussion as follows. 
We start in Section 3.1 with a brief review of
the Planck scale cosmology presented phenomenologically
in Refs.~\cite{bon-reut1,bon-reut2}. 
Our results in
Ref.~\cite{bw4} for the dimensionless gravitational and cosmological constants
at the UV fixed point are reviewed in Section 3.2. In Section 3.3, we 
review the use~\cite{darkuni} our results in Section 3.2 in the context of
the Planck scale cosmology 
scenario in Refs.~\cite{bon-reut1,bon-reut2} to estimate 
the observed value of 
the cosmological constant $\Lambda$. We  
review the use the attendant estimate to constrain
SUSY GUTs. We also address consistency checks on the analysis.
\par
\subsection{\bf Planck Scale Cosmology: A Brief Review}
We begin with the Einstein-Hilbert 
theory
\begin{equation}
{\cal L}(x) = \frac{1}{2\kappa^2}\sqrt{-g}\left( R -2\Lambda\right).
\label{lgwrld1a}
\end{equation} 
Here, $R$ is the curvature scalar, $g$ is the determinant of the metric
of space-time $g_{\mu\nu}$, $\Lambda$ is the cosmological
constant and $\kappa=\sqrt{8\pi G_N}$ for Newton's constant
$G_N$. Employing the phenomenological exact renormalization group
for the Wilsonian~\cite{wil1,wil2,pol1,kgw1,kgw2,kgw3} coarse grained effective 
average action in field space, the authors in Ref.~\cite{bon-reut1,bon-reut2}
have argued that
the attendant running Newton constant $G_N(k)$ and running 
cosmological constant
$\Lambda(k)$ approach UV fixed points as $k$ goes to infinity
in the deep Euclidean regime. This means that 
$k^2G_N(k)\rightarrow g_*,\; \Lambda(k)\rightarrow \lambda_*k^2$
for $k\rightarrow \infty$ in the Euclidean regime.\par
To make contact with cosmology, one may use a connection between 
the momentum scale $k$ characterizing the coarseness
of the Wilsonian graininess of the average effective action and the
cosmological time $t$. The authors
in Refs.~\cite{bon-reut1,bon-reut2} use a phenomenological realization of this latter connection, specifically $k(t)=\frac{\xi}{t}$
for some positive constant $\xi$ determined
from constraints on
physically observable predictions, to
show that the standard cosmological
equations admit of the following extension:
\begin{align}
(\frac{\dot{a}}{a})^2+\frac{K}{a^2}&=\frac{1}{3}\Lambda+\frac{8\pi}{3}G_N\rho,\cr
\dot{\rho}+3(1+\omega)\frac{\dot{a}}{a}\rho&=0,\;\cr
\dot{\Lambda}+8\pi\rho\dot{G_N}&=0,\;\cr
G_N(t)=G_N(k(t)),&\;
\Lambda(t)=\Lambda(k(t)).\cr
\label{coseqn1}
\end{align}
Here, we use a standard notation for the density $\rho$ and scale factor $a(t)$
with the Robertson-Walker metric representation given as
\begin{equation}
ds^2=dt^2-a(t)^2\left(\frac{dr^2}{1-Kr^2}+r^2(d\theta^2+\sin^2\theta d\phi^2)\right)
\label{metric1}
\end{equation}
where $K=0,1,-1$ correspond respectively to flat, spherical and
pseudo-spherical 3-spaces for constant time t.  
For the equation of state we take  
$ 
p(t)=\omega \rho(t),
$
where $p$ is the pressure.
\par
From the UV fixed points for $k^2G_N(k)\equiv g_*$ and
$\Lambda(k)/k^2\equiv \lambda_*$ obtained from their phenomenological, exact renormalization
group (asymptotic safety) 
analysis, the authors in Refs.~\cite{bon-reut1,bon-reut2}
show that the system given above admits, for $K=0$,
a solution in the Planck regime where $0\le t\le t_{\text{class}}$, with
$t_{\text{class}}$ a ``few'' times the Planck time $t_{Pl}$, which joins
smoothly onto a solution in the classical regime, $t>t_{\text{class}}$,
which coincides with standard Friedmann-Robertson-Walker phenomenology
but with the horizon, flatness, scale free Harrison-Zeldovich spectrum,
and entropy problems all solved. The solutions are
achieved purely by Planck scale quantum physics.\par
The phenomenological nature of the analyses in Refs.~\cite{bon-reut1,bon-reut2} 
is made manifest by
the dependencies of
the fixed-point results $g_*,\lambda_*$ on the cut-offs
used in the Wilsonian coarse-graining procedure, for example. We note that 
the key properties of $g_*,\; \lambda_*$ used for these analyses 
are that the two UV limits are both positive and that the product 
$g_*\lambda_*$ is only mildly cut-off/threshold function dependent.
With this latter observations in mind,
we review next the predictions in Refs.~\cite{bw4} for these
UV limits as implied by resummed quantum gravity(RQG) theory~\cite{bw1,bw2,bw5}
and show how to use them to predict~\cite{darkuni} the current value of $\Lambda$. We start the next subsection
with a brief review of the basic principles of RQG theory. 
\par
\subsection{\bf Recapitulation: $g_*$ and $\lambda_*$ in Resummed Quantum  Gravity}
We start with the prediction for $g_*$ in Refs.~\cite{darkuni,bw1,bw2,bw4}. Given that
the theory we use is not very familiar, we review
the main steps in the calculation.
\par
As the graviton couples to an elementary particle 
in the infrared regime which we shall
resum independently of the particle's spin~\cite{wein-qft}, 
we may use a scalar
field to develop the required calculational framework, which we then extend
to spinning particles straightforwardly.  
We follow Feynman in Refs.~\cite{rpf1,rpf2} 
and start with the Lagrangian density for
the basic scalar-graviton system:{\small
\begin{equation}
\begin{split}
{\cal L}(x) &= -\frac{1}{2\kappa^2} R \sqrt{-g}
            + \frac{1}{2}\left(g^{\mu\nu}\partial_\mu\varphi\partial_\nu\varphi - m_o^2\varphi^2\right)\sqrt{-g}\\
            &= \quad \frac{1}{2}\left\{ h^{\mu\nu,\lambda}\bar h_{\mu\nu,\lambda} - 2\eta^{\mu\mu'}\eta^{\lambda\lambda'}\bar{h}_{\mu_\lambda,\lambda'}\eta^{\sigma\sigma'}\bar{h}_{\mu'\sigma,\sigma'} \right\}\\
            & + \frac{1}{2}\left\{\varphi_{,\mu}\varphi^{,\mu}-m_o^2\varphi^2 \right\} -\kappa {h}^{\mu\nu}\left[\overline{\varphi_{,\mu}\varphi_{,\nu}}+\frac{1}{2}m_o^2\varphi^2\eta_{\mu\nu}\right]\\
            &  - \kappa^2 \left[ \frac{1}{2}h_{\lambda\rho}\bar{h}^{\rho\lambda}\left( \varphi_{,\mu}\varphi^{,\mu} - m_o^2\varphi^2 \right) - 2\eta_{\rho\rho'}h^{\mu\rho}\bar{h}^{\rho'\nu}\varphi_{,\mu}\varphi_{,\nu}\right] + \cdots \\
\end{split}
\label{eq1-1}
\end{equation}}
Here,
$\varphi(x)$ can be identified as the physical BEH~\cite{ebh1,ebh2,ebh3,ebh4,atlas-cms-2012a,atlas-cms-2012b,atlas-cms-2012c,atlas-cms-2012d} field as
our representative scalar field for matter,
$\varphi(x)_{,\mu}\equiv \partial_\mu\varphi(x)$,
and $g_{\mu\nu}(x)=\eta_{\mu\nu}+2\kappa h_{\mu\nu}(x)$
where we follow Feynman and expand about Minkowski space
so that $\eta_{\mu\nu}=diag\{1,-1,-1,-1\}$.
We have introduced Feynman's notation
$\bar y_{\mu\nu}\equiv \frac{1}{2}\left(y_{\mu\nu}+y_{\nu\mu}-\eta_{\mu\nu}{y_\rho}^\rho\right)$ for any tensor $y_{\mu\nu}$\footnote{Our conventions for raising and lowering indices in the 
second line of (\ref{eq1-1}) are the same as those
in Ref.~\cite{rpf2}.}.
The bare(renormalized) scalar boson mass here is $m_o$($m$) 
and we set presently the small
observed~\cite{cosm1a,cosm1b,pdg2008} value of the cosmological constant
to zero so that our quantum graviton, $h_{\mu\nu}$, has zero rest mass.
We return to the latter point, however, when we discuss phenomenology.
Feynman~\cite{rpf1,rpf2} has essentially worked out the Feynman rules for (\ref{eq1-1}), including the rule for the famous
Feynman-Faddeev-Popov~\cite{rpf1,ffp1a,ffp1b} ghost contribution required 
for unitarity with the fixing of the gauge
(we use the gauge of Feynman in Ref.~\cite{rpf1},
$\partial^\mu \bar h_{\nu\mu}=0$).
For this material we refer to Refs.~\cite{rpf1,rpf2}. 
We turn now directly to the quantum loop corrections
in the theory in (\ref{eq1-1}).
\par
\begin{figure}
\begin{center}
\epsfig{file=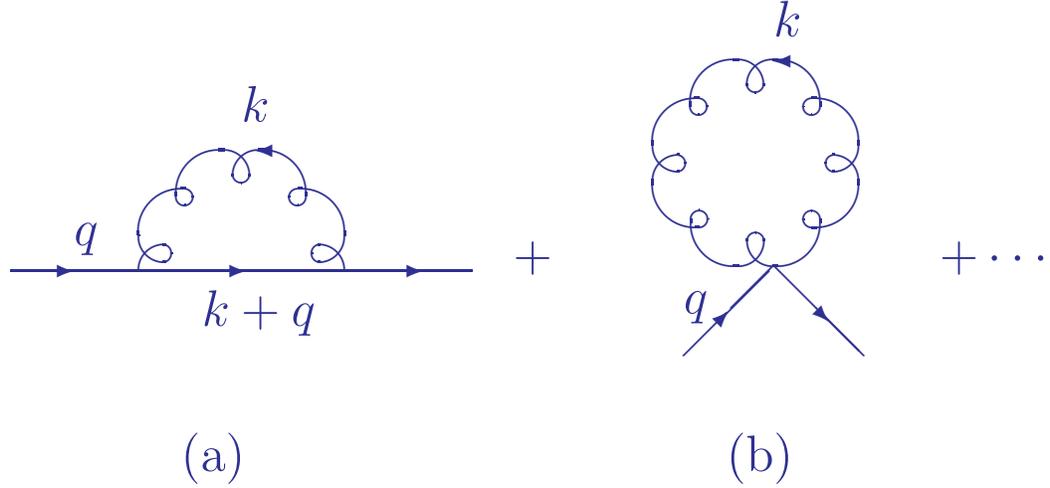,width=140mm}
\end{center}
\caption{\baselineskip=7mm     Graviton loop contributions to the
scalar propagator. $q$ is the 4-momentum of the scalar.}
\label{fig1}
\end{figure}

Referring to Fig.~\ref{fig1}, 
we have shown in Refs.~\cite{bw1,bw2,bw5} that the large virtual IR effects
in the respective loop integrals for 
the scalar propagator in quantum general relativity 
can be resummed to the {\em exact} result
$
i\Delta'_F(k)=\frac{i}{k^2-m^2-\Sigma_s(k)+i\epsilon}
=  \frac{ie^{B''_g(k)}}{k^2-m^2-\Sigma'_s+i\epsilon}
\equiv i\Delta'_F(k)|_{\text{resummed}}
$
for{\small ~~~($\Delta =k^2 - m^2$)
\begin{equation}
\begin{split} 
B''_g(k)&= -2i\kappa^2k^4\frac{\int d^4\ell}{16\pi^4}\frac{1}{\ell^2-\lambda^2+i\epsilon}\\
&\qquad\frac{1}{(\ell^2+2\ell k+\Delta +i\epsilon)^2}\\
&=\frac{\kappa^2|k^2|}{8\pi^2}\ln\left(\frac{m^2}{m^2+|k^2|}\right),       
\end{split}
\label{yfs1} 
\end{equation}}
where the latter form holds for the UV(deep Euclidean) regime, 
so that $\Delta'_F(k)|_{\text{resummed}}$ 
falls faster than any power of $|k^2|$ -- by Wick rotation, the identification
$-|k^2|\equiv k^2$ in the deep Euclidean regime gives 
immediate analytic continuation to the result in the last line of (\ref{yfs1})
when the usual $-i\epsilon,\; \epsilon\downarrow 0,$ is appended to $m^2$. An analogous result~\cite{bw5,darkuni} holds
for m=0. Here, $-i\Sigma_s(k)$ is the 1PI scalar self-energy function
so that $i\Delta'_F(k)$ is the exact scalar propagator. As $\Sigma'_s$ starts in ${\cal O}(\kappa^2)$,
we may drop it in calculating one-loop effects. 
When the respective analogs of $i\Delta'_F(k)|_{\text{resummed}}$\footnote{These follow from
the observation~\cite{bw5,wein-qft} that the IR limit of the coupling of the graviton to a particle is independent of its spin.} are used for the
elementary particles, one-loop 
corrections are finite. In fact, the use of
our resummed propagators renders all quantum 
gravity loops UV finite~\cite{bw1,bw2,bw5,darkuni}. It is this attendant representation
of the quantum theory of general relativity 
that we have called resummed quantum gravity (RQG).
\par
Indeed,
when we use our resummed propagator results, 
as extended to all the particles
in the SM Lagrangian and to the graviton itself, working now with the
complete theory
$
{\cal L}(x) = \frac{1}{2\kappa^2}\sqrt{-g} \left(R-2\Lambda\right)
            + \sqrt{-g} L^{\cal G}_{SM}(x)
$
where $L^{\cal G}_{SM}(x)$ is SM Lagrangian written in diffeomorphism
invariant form as explained in Refs.~\cite{bw2,bw5}, we show in the Refs.~\cite{bw1,bw2,bw5} that the denominator for the propagation of transverse-traceless
modes of the graviton becomes ($M_{Pl}$ is the Planck mass)
$
q^2+\Sigma^T(q^2)+i\epsilon\cong q^2-q^4\frac{c_{2,eff}}{360\pi M_{Pl}^2},
$
where we have defined
$
c_{2,eff}=\sum_{\text{SM particles j}}n_jI_2(\lambda_c(j))
         \cong 2.56\times 10^4
$
with $I_2$ defined~\cite{bw1,bw2,bw5}
by
$
I_2(\lambda_c) =\int^{\infty}_0dx x^3(1+x)^{-4-\lambda_c x}
$
and with $\lambda_c(j)=\frac{2m_j^2}{\pi M_{Pl}^2}$ and~\cite{bw1,bw2,bw5}
$n_j$ equal to the number of effective degrees of particle $j$. 
The details of the derivation of the
numerical value of $c_{2,eff}$ are given in Refs.~\cite{bw5}.
These results allow us to identify (we use $G_N$ for $G_N(0)$) 
$
G_N(k)=G_N/(1+\frac{c_{2,eff}k^2}{360\pi M_{Pl}^2})
$
and to compute the UV limit $g_*$ as
$
g_*=\lim_{k^2\rightarrow \infty}k^2G_N(k^2)=\frac{360\pi}{c_{2,eff}}\cong 0.0442.$
\par
For the prediction for $\lambda_*$, we use the Euler-Lagrange
equations to get Einstein's equation as 
\begin{equation}
G_{\mu\nu}+\Lambda g_{\mu\nu}=-\kappa^2 T_{\mu\nu}
\label{eineq1}
\end{equation}
in a standard notation where $G_{\mu\nu}=R_{\mu\nu}-\frac{1}{2}Rg_{\mu\nu}$,
$R_{\mu\nu}$ is the contracted Riemann tensor, and
$T_{\mu\nu}$ is the energy-momentum tensor. Working then with
the representation $g_{\mu\nu}=\eta_{\mu\nu}+2\kappa h_{\mu\nu}$
for the flat Minkowski metric $\eta_{\mu\nu}=\text{diag}(1,-1,-1,-1)$
we see that to isolate $\Lambda$ in Einstein's 
equation (\ref{eineq1}) we may evaluate
its VEV(vacuum expectation value of both sides). 
On doing this as described in Ref.~\cite{darkuni}, we see that 
a scalar makes the contribution to $\Lambda$ given by\footnote{We note the
use here in the integrand of $2k_0^2$ rather than the $2(\vec{k}^2+m^2)$ in Ref.~\cite{bw4}, to be
consistent with $\omega=-1$~\cite{zeld} for the vacuum stress-energy tensor.}
\begin{equation}
\begin{split}
\Lambda_s&=-8\pi G_N\frac{\int d^4k}{2(2\pi)^4}\frac{(2k_0^2)e^{-\lambda_c(k^2/(2m^2))\ln(k^2/m^2+1)}}{k^2+m^2}\cr
&\cong -8\pi G_N[\frac{1}{G_N^{2}64\rho^2}],
\end{split}
\label{lambscalar}
\end{equation} 
where $\rho=\ln\frac{2}{\lambda_c}$ and we have used the calculus
of Refs.~\cite{bw1,bw2,bw5}.
The standard methods~\cite{darkuni} 
then show that a Dirac fermion contributes $-4$ times $\Lambda_s$ to
$\Lambda$, so that the deep UV limit of $\Lambda$ then becomes, allowing $G_N(k)$
to run,
$\Lambda(k) \operatornamewithlimits{\longrightarrow}_{k^2\rightarrow \infty} k^2\lambda_*,\;
\lambda_* =-\frac{c_{2,eff}}{2880}\sum_{j}(-1)^{F_j}n_j/\rho_j^2
\cong 0.0817$
where $F_j$ is the fermion number of $j$, $n_j$ is the effective
number of degrees of freedom of $j$ and $\rho_j=\rho(\lambda_c(m_j))$.
We note that
$\lambda_*$ would vanish
in an exactly supersymmetric theory.\par
For reference, the UV fixed-point calculated here, 
$(g_*,\lambda_*)\cong (0.0442,0.0817)$, can be compared with the estimates
$(g_*,\lambda_*)\approx (0.27,0.36)$
in Refs.~\cite{bon-reut1,bon-reut2}.
In making this comparison, one must keep in mind that 
the analysis in Refs.~\cite{bon-reut1,bon-reut2} did not include
the specific SM matter action and that there is definitely cut-off function
sensitivity to the results in the latter analyses. What is important
is that the qualitative results that $g_*$ and $\lambda_*$ are 
both positive and are less than 1 in size 
are true of our results as well.
See Refs.~\cite{bw5,darkuni} for further discussion of the relationship between
our $\{g_*,\;\lambda_*\}$ predictions and those in Refs.~\cite{bon-reut1,bon-reut2}.
\par
\subsection{\bf Review of an Estimate of $\Lambda$ and Its Implications}
 
When taken together with those in Refs.~\cite{bon-reut1,bon-reut2}, the results 
reviewed here allow us to estimate the value of $\Lambda$ today. 
To this end, we take the normal-ordered form of Einstein's equation 
\begin{equation}
:G_{\mu\nu}:+\Lambda :g_{\mu\nu}:=-\kappa^2 :T_{\mu\nu}: .
\label{eineq2}
\end{equation}
The coherent state representation of the thermal density matrix then gives
the Einstein equation in the form of thermally averaged quantities with
$\Lambda$ given by our result in (\ref{lambscalar}) summed over 
the degrees of freedom as specified above in lowest order. In Ref.~\cite{bon-reut2}, it is argued that the Planck scale cosmology description of inflation gives the transition time between the Planck regime and the classical Friedmann-Robertson-Walker(FRW) regime as $t_{tr}\sim 25 t_{Pl}$. (We discuss in Ref.~\cite{darkuni} the uncertainty of this choice of $t_{tr}$.)
We thus start with the quantity
$\rho_\Lambda(t_{tr}) \equiv\frac{\Lambda(t_{tr})}{8\pi G_N(t_{tr})}
         =\frac{-M_{Pl}^4(k_{tr})}{64}\sum_j\frac{(-1)^Fn_j}{\rho_j^2}$
and employ the arguments in Refs.~\cite{branch-zap} ($t_{eq}$ is the time of matter-radiation equality) to get the 
first principles field theoretic estimate
\begin{equation}
\begin{split}
\rho_\Lambda(t_0)&\cong \frac{-M_{Pl}^4(1+c_{2,eff}k_{tr}^2/(360\pi M_{Pl}^2))^2}{64}\sum_j\frac{(-1)^Fn_j}{\rho_j^2}\cr
          &\qquad\quad \times \frac{t_{tr}^2}{t_{eq}^2} \times (\frac{t_{eq}^{2/3}}{t_0^{2/3}})^3\cr
    &\cong \frac{-M_{Pl}^2(1.0362)^2(-9.194\times 10^{-3})}{64}\frac{(25)^2}{t_0^2}\cr
   &\cong (2.4\times 10^{-3}eV)^4.
\end{split}
\label{eq-rho-expt}
\end{equation}
where we take the age of the universe to be $t_0\cong 13.7\times 10^9$ yrs. 
In the latter estimate, the first factor in the second line comes from the period from
$t_{tr}$ to $t_{eq}$ which is radiation dominated and the second factor
comes from the period from $t_{eq}$ to $t_0$ which is matter dominated
\footnote{The method of the operator field forces the vacuum energies to follow the same scaling as the non-vacuum excitations.}.
This estimate should be compared with the experimental result~\cite{pdg2008}\footnote{See also Ref.~\cite{sola2} for an analysis that suggests 
a value for $\rho_\Lambda(t_0)$ that is qualitatively similar to this experimental result.} 
$\rho_\Lambda(t_0)|_{\text{expt}}\cong ((2.37\pm 0.05)\times 10^{-3}eV)^4$. 
\par
To sum up, we believe our estimate 
of $\rho_\Lambda(t_0)$
represents some amount of progress in
the long effort to understand its observed value  
in quantum field theory. Evidently, the estimate is not a precision prediction,
as hitherto unseen degrees of freedom, such as those in 
a high scale GUT theory, 
may exist that have not been included in the calculation.\par
Indeed, what would happen to our estimate if there were a GUT theory at high scale? In Ref.~\cite{darkuni} we consider
the susy SO(10) GUT model in Ref.~\cite{ravi-1}
to illustrate how such theory might affect our estimate of $\Lambda$.
In this model, the break-down of the GUT gauge symmetry to the 
low energy gauge symmetry occurs with an intermediate stage with gauge group
$SU_{2L}\times SU_{2R}\times U_1\times SU(3)^c$ where the final break-down to the Standard Model~\cite{gsw1,gsw2,gsw3,gsw4,gsw5,gsw6,gsw7,qcd1,qcd2,qcd3} gauge group, $SU_{2L}\times U_1\times SU(3)^c$, occurs at a scale $M_R\gtrsim 2TeV$ while the breakdown of global susy occurs at the (EW) scale $M_S$ which satisfies $M_R > M_S$. What we find
is that adding the contributions from the new degrees of freedom
for a still viable mass spectrum in this scenario results in a value for 
the RHS of (\ref{eq-rho-expt}) that has the wrong sign with a significance of
many standard deviations. We show~\cite{darkuni} that one can resolve this apparent discrepancy either by adding new particles to the scenario, where the known quarks and leptons are doubled with susy partners for the new families that are lighter
that the families themselves or by moving the mass of the gravitino to a point 
near the GUT scale itself, which is $\sim 4\times 10^{16} GeV $~\cite{ravi-1}. Our result for $\Lambda$ already puts constraints on a class of susy GUT's.\par 
\par 
As we explain in Ref.~\cite{darkuni},
we stress that we actually do not know the precise value of $t_{tr}$ at this point to better than a couple of orders of magnitude which translate to an uncertainty at the level of $10^4$ on
our estimate of $\rho_\Lambda$. We ask the reader to keep this in mind.
\par
We have not mentioned the effect of the various spontaneous symmetry vacuum energies on our 
$\rho_{\Lambda}$ estimate. From the standard methods we know for example that the energy of the broken vacuum for the EW case contributes an amount of order $M_W^4$ to $\rho_\Lambda$. If we consider the GUT 
symmetry breaking we expect an analogous contribution from spontaneous symmetry breaking of order $M_{GUT}^4$. When compared to the RHS of 
our equation for $\rho_\Lambda(t_{tr})$, which is $\sim (-(1.0362)^2W_\rho/64)M_{Pl}^4\simeq \frac{10^{-2}}{64}M_{Pl}^4$, we see that adding these effects thereto would make relative changes in 
our results at the level of 
$\frac{64}{10^{-2}}\frac{M_W^4}{M_{Pl}^4}\cong 1\times 10^{-65} $ and $\frac{64}{10^{-2}}\frac{M_{GUT}^4}{M_{Pl}^4}\cong 7\times 10^{-7}$, respectively, where we use our value of $M_{GUT}$ given above in the latter 
evaluation for definiteness. 
Such small effects are ignored here.
\par
Concerning the impact of our approach to $\Lambda$
on the phenomenology of big bang nucleosynthesis(BBN)~\cite{bbn}, we recall that the authors in Ref.~\cite{bon-reut2}
have already noted that when on passes from the Planck era to the FRW era,
a gauge transformation (from the attendant diffeomorphism invariance) is necessary to maintain consistency with
the solutions of the system (\ref{coseqn1})(or of its more general form as give below) at the boundary between the two regimes at the transition time $t_{tr}$. Requiring that the Hubble parameter be continuous at $t_{tr}$ 
the authors in Ref.~\cite{bon-reut2} arrive at the gauge transformation on the
time for the FRW era relative to the Planck era $t\rightarrow t'=t-t_{as}$
so that continuity of the Hubble parameter at the boundary gives
$\frac{\alpha}{t_{tr}}=\frac{1}{2(t_{tr}-t_{as})}$
when $a(t)\propto t^\alpha$ in the (sub-)Planck regime. This implies $t_{as}=(1-\frac{1}{2\alpha})t_{tr}.$ In our case , we have from Ref.~\cite{bon-reut2} the generic case $\alpha=25$, so that $t_{as}=0.98t_{tr}.$ Here, we use the diffeomorphism invariance of the theory to choose another coordinate transformation for the FRW era, namely, $t\rightarrow t'=\gamma t$ 
as a part of a dilatation
where $\gamma$ now satisfies the boundary condition required for continuity of the Hubble parameter at $t_{tr}$:
$\frac{\alpha}{t_{tr}}=\frac{1}{2\gamma t_{tr}}$  
so that $\gamma=\frac{1}{2\alpha}.$ The model in Ref.~\cite{bon-reut2} purports that, for $t>t_{tr}$, one has the time $t'$ and an effective FRW cosmology with such a small value of $\Lambda$ that it may be treated as zero. Here, we extend this by retaining $\Lambda\ne 0$ so that we may estimate its value. But, with our 
diffeomorphism transformation between the (sub-)Planck regime and the FRW regime, we can see that, at the time of BBN, the ratio of $\rho_\Lambda$ to
$\frac{3H^2}{8\pi G_N}$ is
\begin{equation} 
\begin{split} \Omega_\Lambda(t_{BBN}) &= \frac{M_{Pl}^2(1.0362)^29.194\times 10^{-3}(25)^2/(64 t_{BBN}^2)}{(3/(8\pi G_N))(1/(2\gamma t_{BBN})^2)}\cr &\cong \frac{\pi 10^{-2}}{24}\cr
&= 1.31\times 10^{-3}.\end{split}\label{bbneq1}\end{equation} Thus, at $t_{BBN}$ our $\rho_\Lambda$ is small enough that it has a negligible effect on the standard BBN phenomenology. 
In contrast to what happens in (\ref{eq-rho-expt}), the uncertainty in the value of $\alpha$ does not affect the estimate in (\ref{bbneq1}) because the factors of $\alpha^2=25^2$ cancel between the numerator and the denominator on the RHS in the first line of (\ref{bbneq1}).\par
Turning next to the issue of the covariance of the theory when $\Lambda$ and $G_N$ depend on time, as we explain in Ref.~\cite{darkuni},
we follow in Eqs.(\ref{coseqn1}) the corresponding realization of the improved Friedmann and Einstein equations as given in Eqs.(3.24) in Ref.~\cite{bon-reut1}.  
The more general
realization of (\ref{coseqn1}) is given in Eqs.(2.1) in Ref.~\cite{bon-reut2} -- our discussions in this Section effectively followed the latter realization. The two realizations differ in the solution of the Bianchi identity constraint:
$D^\nu\left(\Lambda g_{\nu\mu}+8\pi G_N T_{\nu\mu}\right)=0;$
for, this identity is solved in (\ref{coseqn1}) for a covariantly conserved
$T_{\mu\nu}$ as well whereas, in Eqs.(2.1) in Ref.~\cite{bon-reut2}, one has the modified conservation requirement
$\dot{\rho}+3\frac{\dot{a}}{a}(1+\omega)\rho= -\frac{\dot{\Lambda}+8\pi\rho \dot{G}_N}{8\pi G_N};$
in (\ref{coseqn1}) the RHS of this latter equation is set to zero. The phenomenology from Ref.~\cite{bon-reut1} is qualitatively unchanged by the simplification in (\ref{coseqn1}) but the attendant details, such as the (sub-)Planck era exponent for the time dependence of $a$, etc., are affected, as is the relation between $\dot{\Lambda}$ and
$\dot{G}_N$ in (\ref{coseqn1}). We note that (\ref{coseqn1}) contains a special case of the more general realization of the Bianchi identity requirement when both $\Lambda$ and $G_N$ depend on time and in this Section we use 
that more general realization. We also note that only when $\dot{\Lambda}+8\pi\rho \dot{G}_N=0$ holds is covariant conservation of matter in the current universe guaranteed and that either the case with or the case without such guaranteed conservation is possible provided the attendant deviation is small. See Refs.~\cite{bianref1,bianref2,bianref3}
for detailed studies
of such deviation, including its maximum possible size.\par
We would note again that 
the model Planck scale cosmology of Bonanno and Reuter
which we use needs more work to remove 
the type of uncertainties which we just elaborated in our estimate of $\Lambda$.
\par
\section{Conclusions}  
We conclude that the standard methods of the operator field do not support
the arguments in Ref.~\cite{attk1} when they are used to argue that
$\Lambda$ and $V_{\text{vac}}$ do not run. The arguments in Ref.~\cite{attk1}
regarding the renormalization invariance of $\Lambda_{\text{phys}}$ and 
$V_{\text{vac-phys}}$,
defined appropriately,
are of course correct.\par
When the Wilsonian running of the parameters in the Einstein-Hilbert theory is taken into account~\cite{bon-reut1,bon-reut2,bon-reut3,bon-reut4,bon-reut5,bon-reut6,bon-reut7a,bon-reut7b,bon-reut7c,bon-reut7d,bon-reut8,bon-reut9,darkuni}, we have shown that, in our resummed quantum gravity realization, we are able to estimate the current value of $\Lambda$. That our result is close to observation is at least encouraging.\par 
\section*{Note Added} In Refs.~\cite{terazawa} it is argued that the physical renormalization group invariant quantities $\Lambda_{\text{phys}}$ and 
$V_{\text{vac-phys}}$ may actually vary with time for example. This would 
from our perspective involve dynamics somewhat 
beyond that which we discuss in this paper.  
\section*{Acknowledgments}
%
We thank Profs. L. Alvarez-Gaume and W. Hollik for the support and kind
hospitality of the CERN TH Division and the Werner-Heisenberg-Institut, MPI, Munich, respectively, where a part of this work was done.
We thank Prof. J. Sola for helpful discussion. 
Work partly supported by US DOE grant DE-FG02-05ER41399 and 
by NATO Grant PST.CLG.980342.\par

\end{document}